\documentstyle[eqsecnum,aps]{revtex}

\begin{document}
\draft
\title{
 Eavesdropper's Optimal Information in Variations
 of Bennett-Brassard 1984 Quantum Key Distribution
 in the Coherent Attacks
}
\author{
WonYoung Hwang \cite{email},
Doyeol (David) Ahn \cite{byline},
and Sung Woo Hwang \cite{byline2}
}
\address{Institute of Quantum Information Processing
and Systems,\\
University of Seoul, 90 Jeonnong, Tongdaemoon,
Seoul 130-743, Korea
}
\maketitle
\begin{abstract}
We calculate eavesdropper's optimal
information on raw bits in Bennett-Brassard 1984
quantum key distribution (BB84 QKD) and
six-state scheme in coherent attacks,
using a formula by Lo and Chau
[Science {\bf 283}, 2050 (1999)] with single photon
assumption.
We find that
eavesdropper's optimal information in QKD without public
announcement of bases [Phys. Lett. A {\bf244}(1998), 489]
is the same as that of a corresponding QKD  {\it with}
it in the coherent attack.
We observe a sum-rule concerning each party's information.
\end{abstract}



\section{Introduction}
\label{sec:level1}
Information processing with quantum systems is an 
interesting
field both theoretically and practically. It may
innovate our fundamental conceptions
on our world \cite{deut}.
And it
is superior to its classical counterpart in some cases:
computing with quantum bits (qubits) enables
factoring large numbers \cite{shor,eke3}, which has
remained intractable with classical computers and
algorithms.
In quantum key distribution (QKD) \cite{wies}-\cite{mull},
it is possible for two legitimate users Alice and Bob to
distribute keys with security that quantum
mechanical laws afford.

Security of QKD had been (tentatively) accepted
on the basis of the no-cloning theorem \cite{woot,diek}
and it might be the first practical quantum
information processor \cite{mull}.
However it is only recently that
its unconditional security is proved \cite{maye}-\cite{sho2}.
Since the original work \cite{bene}, more and more
sophisticated attacks have been considered:
intercept-resend strategy with orthogonal measurement
in general bases, attacks with generalized (or positive
operator valued) measurements, and
the most general coherent (collective or joint) attacks
where all
qubits are coherently treated as a whole quantum system
were considered in Refs. \cite{hutt},
\cite{eke2}-\cite{slut}, and
\cite{biha}-\cite{sho2}, respectively.
One of the reasons making the proof complicated is
that there is inevitable residual noise in
real quantum channel. And the
natural noise cannot be discriminated from what
Eavesdropper's (Eve's) tapping on the channel causes.
Thus raw bits must be processed in such a way so that Eve
has essentially zero information about the final
corrected bits. This might be done either quantum
or classical information processing.
In the former case, errors are removed by quantum
error correcting codes or purification protocol
\cite{ben4}. In the latter case, errors are
corrected with certain classical error correcting codes.
In particular, in case of
Ref. \cite{sho2}
classical error correction codes
associated with Calderbank-Shor-Steane
\cite{cald,stea} quantum error correcting codes
are used. The security of the method
against coherent attacks are
proved in Refs. \cite{maye},\cite{bih2,sho2}.
However, before such
elegant proofs were given, Eve's optimal
information on raw bits for various attacks
were estimated \cite{hutt}-\cite{fuc2}
for classical privacy amplification, where Eve's information
is deleted.
Although it is not rigorously
proven such methods are secure against coherent attacks,
it seems to
be so for almost practical purposes and thus the security
of such methods had been widely accepted.
However, the estimations have been confined
to individual
attacks where each qubit is
separately treated  \cite{hutt}-\cite{fuc2}.
Estimating Eve's optimal information in coherent attacks
was in itself an interesting and unsolved problem until
formulas for it are given recently \cite{lo,bih2}.
Thus it is worthwhile to have explicit estimation of it.
In this paper, we calculate
Eve's optimal information about raw bits
in BB84, six-state schemes
in the coherent attack using the formula given by Lo and
Chau \cite{lo}, with single photon (or quantum) assumption.
Then we
consider multiple-basis scheme where a number of
bases are adopted: we find Eve's optimal information in
multiple-basis scheme is the same as that of six-state
scheme. We consider another variation of BB84
scheme, QKD without public announcement of bases
\cite{hwan}. We argue the formula is also valid for
it. We also find
Eve's optimal information in it is the same as that
of a corresponding QKD with public announcement of bases.
We observe that sum of mutual information between Alice 
and Bob, and Eve's information is constant in the case of 
BB84 scheme.
\section{formula for eavesdropper's optimal information}
\label{sec:level1}
Derivation of 
the formula for Eve's optimal information $I_{Eve}$
in Eq. (\ref{h}) is briefly discussed in
Ref. \cite{lo}. In this section, we give a more detailed
derivation of the formula in a self-consistent manner.
The entanglement-based schemes \cite{ben2}
can be reduced to BB84-like scheme \cite{bene}.
Thus $I_{Eve}$ in
entanglement-based scheme which we calculate
is the same as that in BB84-like scheme.
First we introduce entanglement-based scheme.
With the convention of Refs. \cite{ben4,lo},
the Bell basis vectors
$|\Psi^{\pm}\rangle$ ($=|01\rangle \pm |10\rangle$)
and $|\Phi^{\pm}\rangle$
($=|00\rangle \pm |11\rangle$) are represented by
two classical bits,
\footnote{In this paper obvious normalization constants
are omitted. It is noted that one should never think of
the Bell basis vectors as direct product state since
they are maximally entangled.}
\begin{eqnarray}
 \label{a}
 |\Phi^+\rangle = \tilde{0} \tilde{0}, \hspace{5mm}
 |\Psi^+\rangle = \tilde{0} \tilde{1}, \hspace{5mm}
 |\Phi^-\rangle = \tilde{1} \tilde{0}, \hspace{5mm}
 |\Psi^-\rangle = \tilde{1} \tilde{1}.
\end{eqnarray}
Eve is supposed to prepare a state
$|\Psi^-\rangle \otimes |\Psi^-\rangle \otimes
\cdot \cdot \cdot \otimes |\Psi^-\rangle$, if she does it
honestly. But she may
prepare other state most generally
\begin{equation}
 \label{b}
 |u\rangle= \sum_{i_1, i_2, \cdot \cdot \cdot, i_N} \sum_{j}
           \alpha_{i_1, i_2, \cdot \cdot \cdot, i_N, j}
           |i_1, i_2, \cdot \cdot \cdot, i_N\rangle \otimes
           |j\rangle,
\end{equation}
where $i_k$ denotes the state of the $k$-th pair,
which runs from
$\tilde{0} \tilde{0}$ to $\tilde{1} \tilde{1}$,
$\alpha_{i_1, i_2,
\cdot \cdot \cdot, i_N, j}$'s are some complex coefficients,
and the $|j\rangle$ values form an orthonormal basis for
the ancilla.
Eve gives this state
to Alice and Bob, the two legitimate
participants who will secretly exchange messages.
On each particle, they independently and randomly performs
measurements among $\hat{S}_z$
(orthogonal measurement composed of two projection operators
$|0\rangle \langle0|$ and $|1\rangle \langle1|$),
$\hat{S}_x$ (that of $|\bar{0}\rangle \langle \bar{0}|$
and $|\bar{1}\rangle \langle\bar{1}|$), and
$\hat{S}_y$ (that of
$|\bar{\bar{0}}\rangle \langle \bar{\bar{0}}|$
and $|\bar{\bar{1}}\rangle \langle \bar{\bar{1}}|$),
where
$|\bar{0}\rangle=|0\rangle+|1\rangle$,
$|\bar{1}\rangle=|0\rangle-|1\rangle$
and
$|\bar{\bar{0}}\rangle=|0\rangle+i|1\rangle$,
$|\bar{\bar{1}}\rangle=|0\rangle-i|1\rangle$.
Then
Alice and Bob compare their bases by public discussion
and they discard their data of the case where
the bases are not matched.
Then Alice and Bob publicly announce
some randomly chosen subsets of remaining data.
They count the number $N_{para}$ of the case where
the results are the same and the number $N_{anti}$ of
the case where the results are opposite.
Alice and Bob calculate Eve's optimal
information $I_{Eve}$ about their results as a function
of error rate $D\equiv N_{para}/(N_{para}+N_{anti})$.
When $D$ is too high, they abort the protocol and restart
it.
Otherwise, they process the raw bits (the data)
into final key about which Eve has essentially zero
information. This completes a description on
entanglement-based scheme.

Now, let us consider Eve's optimal information.
Roughly speaking, the higher the error rate $D$ becomes,
the more states other than $\Psi^-$ are contained in the
state of Eq. (\ref{b}).  Thus 
the entropy of reduced density operator 
of Alice and Bob's qubits become higher and 
Eve can extract more
information on the bits (see Eq. (\ref{g})).
More precisely,
first note the following equivalences.
\begin{eqnarray}
\label{c}
|00\rangle \langle00|+|11\rangle \langle11|&=&
|\Phi^+\rangle \langle \Phi^+|+|\Phi^-\rangle \langle \Phi^-|
,\nonumber\\
|\bar{0}\bar{0}\rangle \langle\bar{0}\bar{0}|+
|\bar{1}\bar{1}\rangle \langle\bar{1}\bar{1}| &=&
|\Phi^+\rangle \langle \Phi^+|+|\Psi^+\rangle \langle \Psi^+|
,\nonumber\\
|\bar{\bar{0}}\bar{\bar{0}}\rangle
 \langle \bar{\bar{0}} \bar{\bar{0}}|+
|\bar{\bar{1}}\bar{\bar{1}}\rangle
 \langle \bar{\bar{1}} \bar{\bar{1}}|  &=&
|\Phi^-\rangle \langle \Phi^-|+|\Psi^+\rangle \langle \Psi^+|
.
\end{eqnarray}
This means that the error rate $D$ that Alice and Bob estimate
from their measurements on qubits in $z,x$, and $y$
basis are the same as they would have estimated using the
Bell basis measurement \cite{sho2}.
So we may estimate $D$ using the
Bell basis measurement. Let us consider the state.
Assume that Eve had performed Bell basis measurement on
all qubits in the state and then sent them to Alice and Bob.
\footnote{One should not regard Eve as doing the
measurement in the real protocol.
It is here a hypothetical one for making the
estimation easier. 
Alice and Bob's measurement is real and thus perturbs the 
state. However, it does not matter here because their 
measurement is local process which does not increase 
entanglement between them and Eve.}
Then Alice and Bob perform Bell basis measurement on some
subsets of the qubits, according to the
scheme. After Eve did the pre-measurement, the state
reduces to a mixed state
\begin{eqnarray}
 \label{d}
 \rho &=& \sum_{i_1, i_2, \cdot \cdot \cdot, i_N}
       P_{i_1, i_2, \cdot \cdot \cdot, i_N}
         |i_1, i_2, \cdot \cdot \cdot, i_N \rangle
 \langle i_1, i_2, \cdot \cdot \cdot, i_N|,
 \\
\mbox{where} &&  \hspace{5mm}
 \label{e} 
 P_{i_1, i_2, \cdot \cdot \cdot, i_N}=
 \sum_{j}|\alpha_{i_1, i_2, \cdot \cdot \cdot, i_N, j}|^2.
\end{eqnarray}
However, in this case Eve's and [Alice+Bob]'s
measurements have common eigenvectors (the Bell basis),
and thus Eve's
pre-measurement do not change statistics of [Alice+Bob]'s
later measurement.
So we may do our estimation of $D$
for the mixed state in Eq. (\ref{d}) instead of
Eq. (\ref{b}).
Then $D$ for the state is given by the following.
\begin{equation}
 \label{f}
 D= \sum_{i_1, i_2, \cdot \cdot \cdot, i_N}
       P_{i_1, i_2, \cdot \cdot \cdot, i_N}
        D(i_1, i_2, \cdot \cdot \cdot, i_N),
\end{equation}
where $D(i_1, i_2, \cdot \cdot \cdot, i_N)$ is
error rate that the state
$|i_1, i_2, \cdot \cdot \cdot,i_N \rangle$ induces.
$D(i_1,i_2, \cdot \cdot \cdot, i_N)$ depends on the
way of checking errors in schemes and will be calculated
in next section. Using Eq. (\ref{f}), we can
calculate expected error rate $D$ for a state with
a certain
$\alpha_{i_1, i_2,\cdot \cdot \cdot, i_N, j}$'s.

On the other hand,
\begin{equation}
 \label{g}
 I_{Eve} \leq S(\rho_{AB}),
\end{equation}
where $S$ is the von Neumann entropy and $\rho_{AB}=$
Tr$_{Eve}|u\rangle \langle u|$
(see {\it Lemma 2}$\hspace{1mm}$ of
the supplementary material of Ref. \cite{lo}).
There are numerous sets of
$\alpha_{i_1, i_2,\cdot \cdot \cdot, i_N, j}$
that give rise to a certain error rate $D$.
What Eve has to do is maximizing her information for a
certain error rate $D$. Thus she has to choose one among
the sets of $\alpha_{i_1, i_2,\cdot \cdot \cdot, i_N, j}$
which give maximal entropy. By inspection, we can see the
maximal entropy is obtained when
all $|i_1, i_2, \cdot \cdot \cdot, i_N \rangle$ giving
the error rate $D$ are prepared with equal probability
$P_{i_1, i_2, \cdot \cdot \cdot, i_N}$.
Then we have
\begin{eqnarray}
 \label{h}
 I_{Eve} \leq -\sum_{i_1, i_2, \cdot \cdot \cdot, i_N}
                   P_{i_1, i_2, \cdot \cdot \cdot, i_N}
              \log \hspace{1mm}
                   P_{i_1, i_2, \cdot \cdot \cdot, i_N}
        \hspace{5mm}
        = \log \hspace{1mm} \Omega,
\end{eqnarray}
where $\Omega$ is the number of distinct
$|i_1, i_2, \cdot \cdot \cdot, i_N \rangle$s
giving an error rate $D$.
(In this paper $\log \equiv \log_2$.)
\section{Optimal information in BB84, six-state
         and multiple-basis scheme}
\label{sec:level1}
First we calculate $I_{Eve}$
of BB84 scheme:
let us calculate $D(i_1, i_2, \cdot \cdot \cdot, i_N)$
for the scheme, where
Alice and Bob check errors by either
$|00\rangle \langle00|+|11\rangle \langle11|$
 or
$|\bar{0}\bar{0}\rangle \langle\bar{0}\bar{0}|+
|\bar{1}\bar{1}\rangle \langle\bar{1}\bar{1}|$.
So probability that $|\Psi^-\rangle$,
$|\Phi^-\rangle$, $|\Psi^+\rangle$,
and $|\Phi^+\rangle$ are detected in
error checking are $0, 1/2, 1/2$, and $1$,
respectively, by Eq. (\ref{c}). Thus,
\begin{equation}
 \label{i}
 D(i_1, i_2, \cdot \cdot \cdot, i_N)=
 \frac{1}{N}(\frac{b}{2}+ \frac{c}{2}+ d),
\end{equation}
where $a$,$b$,$c$,and $d$ are the number of elements of
the set $A=\{i_k| i_k= \tilde{1}\tilde{1}\}$, $B=\{i_k| i_k=
\tilde{1}\tilde{0}\}$, $C=\{i_k| i_k=\tilde{0}\tilde{1}\}$,
and $D=\{i_k| i_k=\tilde{0}\tilde{0}\}$, respectively
($k$=1,2,...,N).
We note that Eq. (\ref{i}) is statistically satisfied only
when Eve
does not know the encoding bases while she has access to the
qubits: if she knows which pairs of qubits will be chosen
for estimation of the error rate $D$, she can cheat by
sending $\Psi^-$ for all the chosen pairs while sending
one of the four Bell states for other pairs.
In order to give an error rate $D$,
\begin{equation}
 \label{j}
 D(i_1, i_2, \cdot \cdot \cdot, i_N)=
 \frac{1}{N}(\frac{b}{2}+ \frac{c}{2}+ d)=D.
\end{equation}
The number $\Omega$ of  $i_1, i_2, \cdot \cdot \cdot, i_N$s that
satisfies Eq. (\ref{j}) is given by
\begin{equation}
 \Omega=\sum_{\frac{1}{2}(b+c)+d=D}
  \frac{N!}{a!\hspace{1.7mm} b!\hspace{1.7mm} c!\hspace{1.7mm} d!}.
\end{equation}
Among many summed terms,
$\Omega$ is dominated by maximal (typical) one.
Thus we obtain
\begin{equation}
 \label{e}
 \log \hspace{1.5mm} \Omega=\mbox{Max} \hspace{2mm} \{
            -(a \hspace{1mm} \log {\frac{a}{N}}
             +b \hspace{1mm} \log {\frac{b}{N}}
             +c \hspace{1mm} \log {\frac{c}{N}}
             +d \hspace{1mm} \log {\frac{d}{N}}) \}
\end{equation}
By inspection, we can see that the maximum is obtained
when $b=c$. Then with
Eq. (\ref{j}),
\begin{equation}
 \label{m}
 \log \hspace{1.5mm} \Omega= \mbox{Max} \hspace{2mm} \{
              (N-2ND+d)\hspace{1mm} \log {\frac{N-2ND+d}{N}}
              + 2(ND-d)\hspace{1mm} \log {\frac{ND-d}{N}}
              +  d \hspace{1mm} \log {\frac{d}{N}}) \}.
\end{equation}
The maximum is obtained when the term's differential is zero or $d=
ND^2$.
\begin{equation}
 \label{n}
 \log \hspace{1.5mm}
 \Omega= -N \{(1-2D+D^2)\hspace{1mm} \log (1-2D+D^2)
                    +2(D-D^2)\hspace{1mm} \log (D-D^2)
                         +D^2\hspace{1mm} \log  D^2 \}.
\end{equation}
Before comparing it with $I_{Eve}$ for incoherent attacks
(Eq. (65) of Ref. \cite{fuc2}),
our $I_{Eve}$ should be divided by $2N$ since it is
the information about $N$ pairs of particles. Then,
\begin{eqnarray}
 \label{o}
 I_{Eve}&\leq& -\frac{1}{2}
 \{(1-2D+D^2) \hspace{1mm} \log (1-2D+D^2)
               +2(D-D^2)\hspace{1mm} \log (D-D^2)
               +D^2\hspace{1mm} \log  D^2 \},
                                    \nonumber\\
    &=& -[D\hspace{1mm} \log  D+ (1-D)\hspace{1mm} \log (1-D)].
\end{eqnarray}
Eq. (\ref{o}) is plotted in Fig. 1 among others.

Next we
calculate $I_{Eve}$ of the six-state scheme \cite{brus} in the
same way:
in the scheme \cite{brus} one of the three
measurements in Eq. (\ref{c})
is performed with equal probabilities.
So we obtain
\begin{equation}
 \label{p}
 D(i_1, i_2, \cdot \cdot \cdot, i_N)=
 \frac{1}{N}(\frac{2}{3}b+ \frac{2}{3}c+ \frac{2}{3}d).
\end{equation}
In order to give an error rate $D$,
\begin{equation}
 \label{q}
 \frac{1}{N}(\frac{2}{3}b+ \frac{2}{3}c+ \frac{2}{3}d)= D.
\end{equation}
The maximum is obtained when $b=c=d=ND/2$.
Then,
\begin{equation}
 \label{r}
 I_{Eve}\leq -\frac{1}{2}
      \{ (1-\frac{3}{2}D)\hspace{1mm} \log (1-\frac{3}{2}D)
          + \frac{3}{2}D \hspace{1mm} \log  \frac{D}{2} \}.
\end{equation}
As we see in Fig. 1, $I_{Eve}$ of Eq. (\ref{r})
is lower than that of Eq. (\ref{p}),
which means that the six-state scheme is more
advantageous than the BB84 scheme in the case of
coherent attacks, too.

 Now we address the multiple-basis scheme. In the scheme
many bases are adopted while two and three bases are adopted
in the BB84 and six-state scheme, respectively. We assume
the bases are uniformly distributed on the Bloch sphere.
(Schemes with
non-uniform distributions do not seem to be more
advantageous than the uniform one.)
It is shown in Ref. \cite{brus} that the multiple basis
scheme does not give more security than the six-state
scheme within the individual attack. 
So we can expect that this is the case
in the coherent attack. Here we show
that the multiple-basis scheme is indeed
no more advantageous than
the six-state scheme in the coherent attack: let us
compute the average
probability $p$ that a Bell state
induces parallel result when they are measured in one of the
many bases uniformly distributed on the Bloch sphere. We can
easily see that $p(|\Psi^-\rangle)=0$ since $|\Psi^-\rangle$
induces only anti-parallel results for any basis. We can also
see
\begin{equation}
 \label{s}
 p(|\Phi^-\rangle)= \int p(|\Phi^-\rangle,\theta) d\Omega =
            \int^{\pi}_0 \sin^2\theta \frac{\sin\theta}{2}
             d\theta
          = \frac{2}{3},
\end{equation}
where $p(|\Phi^- \rangle,\theta)$
is the probability density that $|\Phi^- \rangle$
induces parallel results for a measurement along a basis that makes
an angle $\theta$ with $z$ axis and $\Omega$ is the solid angle. In
a similar way,
\begin{equation}
 \label{t}
 p(|\Psi^+\rangle)=p(|\Phi^+\rangle)=\frac{2}{3}.
\end{equation}
Thus for the multiple-basis scheme we have the same equation as
Eq. (\ref{q}). Accordingly, $I_{Eve}$ of this scheme is the same as
that of six-state scheme. We can also consider a
multiple-basis
scheme where the bases are uniformly distributed in $z-x$ plane. We
can also show in a similar way that this multiple basis scheme in
the plane is no more advantageous than the BB84 scheme:
\begin{eqnarray}
 \label{u}
 p(|\Psi^-\rangle)=0, 
 \hspace{5mm}
 p(|\Phi^-\rangle)= p(|\Psi^+\rangle)=
 \frac{1}{\pi}\int^{\pi}_0 \sin^2\theta d\theta
            = \frac{1}{2},
 \hspace{5mm}
 p(|\Phi^+\rangle)= 1.
\end{eqnarray}
\section{optimal information in
 QKD without public announcement of bases}
\label{sec:level1}
Here we show that $I_{Eve}$ of
QKD without public announcement of bases \cite{hwan} is the
same as that of
a corresponding one with public announcement of
bases. Let us consider a scheme corresponding to BB84  
\cite{hwan}.
In the scheme,
Eve knows which and which qubits are encoded in the same
basis while she does not know which basis between $z$
and $x$ they
are. In this case the probability that Eve will make
a right guess
of the encoding bases is still 1/2,
which is the same as
that in the case of BB84 scheme.
Thus Eq. (\ref{i}) is also valid and
later procedures for calculation of $I_{Eve}$ are the
same as that
of BB84 scheme. So $I_{Eve}$ of QKD without
public announcement of
bases is the same as that of BB84 scheme.
The idea of
QKD without public announcement of bases are straitforwardly
applied to the six-state scheme.
And we can see that
$I_{Eve}$ of the six-state QKD without public announcement
of bases is the
same as that of the six-state scheme with it.
However, if $I_{Eve}$ of
both schemes are the same, we can say
that QKD without public
announcement of bases is more advantageous than
either BB84 scheme or six-state scheme:
while in either BB84 scheme or six-state scheme
full information about the
encoding bases are given to Eve after the qubits
have arrived at Bob, in QKD without public announcement
of bases only
partial information (which and which are the same basis)
about the encoding bases are given to Eve.
To summarize this section, these facts suggest
that QKD without public announcement of
bases is at least 
as secure as either BB84 or six-state scheme
even in coherent attacks.
\section{discussion and conclusion}
\label{sec:level1}
It is  interesting that the sum of $I_{Eve}$ of BB84 scheme
in Eq. (\ref{o}) and
\begin{equation}
I_{AB}= 1+ D\hspace{1mm} \log D+ (1-D)\hspace{1mm} \log (1-D)
\end{equation}
is constant. That is,
\begin{equation}
 I_{Eve}+ I_{AB}= 1.
\end{equation}
This indicates something is conserved.
Roughly speaking, QKD could be interpreted by the quantum
information conservation:\footnote{Attempts for
establishment of rigorous
quantum information conservation theorems can be found in
Refs. \cite{horo,pope}.}
since the total quantum information that
Alice have sent is conserved, the more quantum information
Eve gets, the less quantum information given to Bob.
It should be noted that the results for $I_{Eve}$ in this
paper are asymptotically valid in the limit when the number
of employed qubits become large.

In conclusion, we have calculated eavesdropper's optimal
information $I_{Eve}$ on raw bits in BB84 and
six-state scheme
in coherent attacks, using the formula (Eq. (\ref{h}))
by Lo and Chau \cite{lo}, assuming single quantums are used.
We have shown that  $I_{Eve}$ in
multiple-basis scheme is the same as that of six-state
scheme. We have considered QKD without public
announcement of bases \cite{hwan}: we found that  $I_{Eve}$
in it
is the same as that of a corresponding QKD {\it with}
public announcement of bases in the coherent attacks.
This fact suggests that QKD without public announcement of
bases is as secure as either
BB84 or six-state scheme in coherent attacks, too.
We observed that $I_{Eve}+ I_{AB}= 1$ in the case of
BB84 scheme.
\acknowledgments
This work was supported by the Korean Ministry of Science
and Technology through the Creative Research Initiatives
Program under Contract No. 99-C-CT-01-C-35.

\vspace{7mm}
FIGURE CAPTION:\\
 the solid line (the upper one): $I_{Eve}$ in BB84 scheme,
                                          Eq. (3.7)\\
 the dotted line (the middle one): $I_{Eve}$ in six-state scheme,
                                   Eq. (3.10)\\
 the dot-dashed line (the lower one): Eq. (65) of Ref. \cite{fuc2}\\
 the dashed line: $I_{AB}$,
                  Eq. (5.1)
\end{document}